\documentclass{article}

\PassOptionsToPackage{comma,square,sort&compress,numbers}{natbib}

    \usepackage[preprint]{neurips_2025}

\usepackage[utf8]{inputenc} 
\usepackage[T1]{fontenc}    
\usepackage[pagebackref,breaklinks,colorlinks]{hyperref}       
\usepackage{url}            
\usepackage{booktabs}       
\usepackage{amsfonts}       
\usepackage{nicefrac}       
\usepackage{microtype}      
\usepackage{xcolor}         

\usepackage{bm}
\usepackage{amsmath}
\usepackage[ruled,noline,linesnumbered]{algorithm2e}
\usepackage{setspace}
\usepackage{extarrows}
\usepackage{graphicx}
\usepackage{multirow}
\usepackage{makecell}

\title{Generative Image Compression by Estimating Gradients of the Rate-variable Feature Distribution}

\author{
  \makecell{\hspace{-9pt}Minghao Han, \hspace{-1pt}
  Weiyi You, \hspace{-1pt}
  Jinhua Zhang, \hspace{-1pt}
  Leheng Zhang, \hspace{-1pt}
  Ce Zhu, \hspace{-1pt}
  Shuhang Gu\thanks{Corresponding Author}}\\
  \hspace{-0.4cm}University of Electronic Science and Technology of China \hspace{0pt}\\
  \tt\small{\{minghao.hmh, shuhanggu\}@gmail.com}
}

\begin{document}

\maketitle

\begin{abstract}
While learned image compression (LIC) focuses on efficient data transmission, generative image compression (GIC) extends this framework by integrating generative modeling to produce photo-realistic reconstructed images. In this paper, we propose a novel diffusion-based generative modeling framework tailored for generative image compression. Unlike prior diffusion-based approaches that indirectly exploit diffusion modeling, we reinterpret the compression process itself as a forward diffusion path governed by stochastic differential equations (SDEs). A reverse neural network is trained to reconstruct images by reversing the compression process directly, without requiring Gaussian noise initialization. This approach achieves smooth rate adjustment and photo-realistic reconstructions with only a minimal number of sampling steps. Extensive experiments on benchmark datasets demonstrate that our method outperforms existing generative image compression approaches across a range of metrics, including perceptual distortion, statistical fidelity, and no-reference quality assessments. The code is available \href{https://github.com/LabShuHangGU/RDM}{here}.
\end{abstract}

\section{Introduction}\label{introduction}

Image compression techniques aim to encode images into the shortest possible bit streams for efficient data transmission. 
Recent studies have developed learned image compression (LIC) methods~\cite{balle2016end,balle2018variational,minnen2018joint,minnen2020channel,cheng2020learned,he2022elic,liu2023learned,han2024causal,lu2025learned,qian2022entroformer} that achieve superior rate-distortion performance compared to conventional codecs~\cite{bpg,vvc}. 
However, the compression process tends to sacrifice the details in images, and the optimization objective, i.e., the rate-distortion loss function, restricts the ability of decoder to restore the lost image details, which finally results in a blurred unrealistic recovered image. 
The pursuit of realism has given rise to a range of generative image compression methods.

Generative image compression (GIC) methods introduce impressive generative modeling techniques, e.g., generative adversarial networks (GANs)~\cite{goodfellow2014generative}, vector-quantized variational autoencoders (VQ-VAEs)~\cite{esser2021taming}, and diffusion-based models~\cite{song2019generative,ho2020denoising}, to obtain the capability of prior distribution modeling, thus improving human-perceptual performance with guaranteed fidelity. 
In their seminal work, perceptual loss, GANs~\cite{agustsson2019generative,mentzer2020high,tschannen2018deep}, and its discriminator variants~\cite{muckley2023improving} are first utilized to finetune the basic image compression network, which allows the decoder in the autoencoder to complement the image details.
For the VQ-based method~\cite{mao2024extreme,jia2024generative,li2024once}, the prior distribution of compressed latent variables is modeled as classification probabilities according to the codebook. 
Compared to GAN-based and VQ-based methods, diffusion modeling decouples their once-through distribution transformations into asymptotic stochastic processes, which significantly enhances generation performance.
The existing diffusion-based methods~\cite{yang2023lossy,hoogeboom2023high,careil2023towards,lei2023text+} can be viewed as a form of ``post-processing'', where diffusion models are employed to enhance the compressed data---an indirect approach aimed at supplementing the lost details. 
However, such an indirect method may not fully harness the potential of diffusion modeling.

\begin{figure}[t]
    \centering
    \includegraphics[width=1\linewidth]{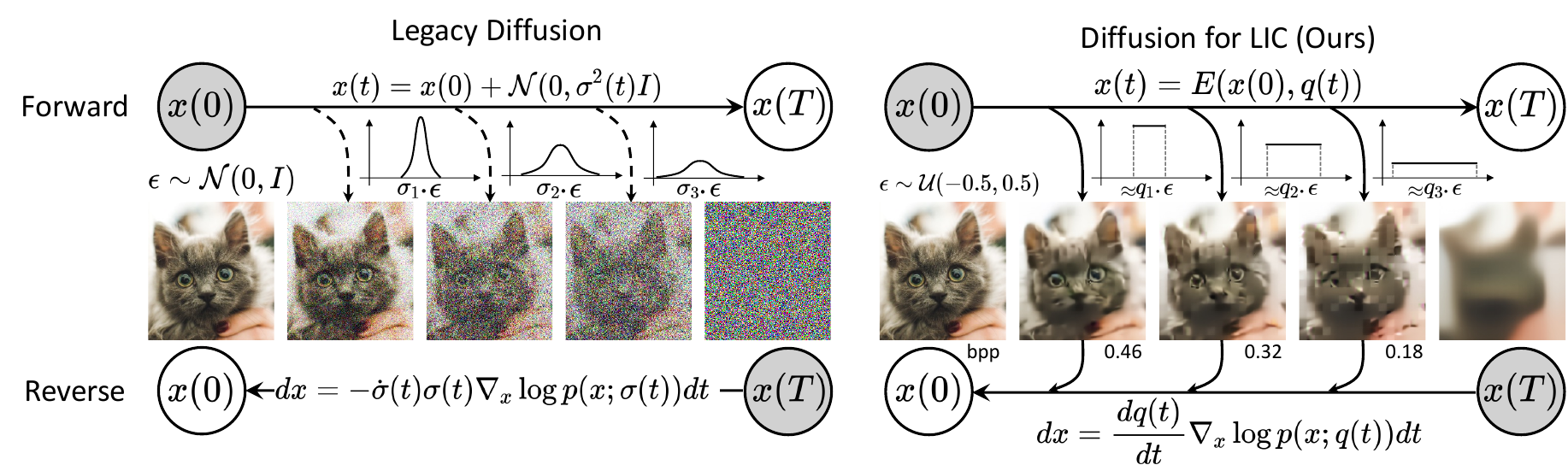}
    \caption{\textbf{Left: }The forward and reverse process of legacy diffusion for comparison. The legacy diffusion consists of transforming data to a simple noise distribution and a reverse ODE to restore the original data. \textbf{Right: The overview pipeline of our method.} The forward process is defined as the entropy model compressing the data. With the bit rates decrease, the compressed images retain less details (please \textit{zoom in} for better visualization). We can reverse such ODE at any intermediate time to recover the data under various compression rates. This makes a full use of the benefits of diffusion modeling and an organic integration of LIC and diffusion.}
    \label{fig:pipeline}
\end{figure}

Revisiting diffusion modeling in the context of image generation, a forward process is constructed by progressively corrupting data with increasing Gaussian noise, and generative modeling is achieved by training a sequence of probabilistic models to reverse the corruption process.
Rate-variable quantization, i.e. quantize feature with different quantization factors, could establish a similar process of gradually distorting high quality data.
As shown in Fig.~\ref{fig:pipeline}, legacy diffusion model adds Gaussian noise to distort the data while the corruption process of rate-variable quantization is often formulated as additive noise with uniform distribution~\cite{balle2016end,balle2018variational}.
Following this, we argue that the rate-variable compression forms a particular forward process with additive noise, and that the goal of restoring the details oriented to the state before compression can be achieved by reversing it.
Taking inspiration from generative modeling through stochastic differential equations (SDEs)~\cite{song2020score}, we describe the aforementioned forward process (compression) and reverse process with the help of SDEs.
That is, it is possible to construct intermediate probability distributions for a stochastic process using a variety of compression rates, and we train a neural network to model the score of a compression-rate-dependent marginal distribution of the training data corrupted by compressing.
Moreover, with the adopted rate-variable quantization parameterizing the compression process into a single quantization factor, this exactly facilitates the formulation of our customized forward and reverse processes. 
The above framework establishes a sequence of distributions for generative modeling, result in a more natural integration of diffusion model and learned image compression, which could support rate-variable generative compression with only a minimal number of reverse steps.

Taken together, we propose generative image compression by estimating
gradients of rate-variable feature distribution, offering a novel perspective beyond conventional paradigms. 
Our method constitutes a customized diffusion framework designed for image compression, featuring both \textbf{flexible rate adjustment} and \textbf{high-fidelity image reconstruction}.
Our rate-variable model is able to outperform current fixed-rate state-of-the-art methods on a range of
perceptual metrics. 
We believe that this work will inspire further innovations in broader areas, particularly in the modification of the diffusion modeling core to suit a variety of specific research scenarios.

\section{Background}\label{background}

In this section, we present the related work and the existing technologies that we adopt in our proposed method. 
First, we review a brief theory of denoising score matching modeling~\cite{song2019generative} as a preliminary for our specialized diffusion models. 
Next, we describe the adopted rate-variable compression strategy for the sake of understanding our proposed framework.
Finally, we introduce recent representative GIC methods that integrate with generative modeling.

\subsection{Denoising Score Matching}

Given the data distribution \(p_{data}(\bm x)\), the idea of score matching is to find the score/gradients of the data distribution \(p_{data}(\bm x)\), that is, the fastest growing direction of the log probability density of the data \(s(\bm x)=\nabla_{\bm x}\log{p(\bm x)}\). 
To obtain the score, Song et al.~\cite{song2019generative} proposed the denoising score matching and noise conditional score networks (NCSN): 
To consider the family of mollified distributions \(p(\bm x;\sigma)\) obtained by adding i.i.d. Gaussian noise of standard deviation \(\sigma\) to the data, the diffusion model sequentially denoises from a pure Gaussian noise \(\bm x_0\sim\mathcal{N}(\textbf{0},\sigma_0^2 \textbf{I})\) into intermediate states \(\bm x_i\sim p(\bm x;\sigma_i)\) with decreasing noise levels \(\sigma_i >\sigma_{i+1}\). 
It ends up with the target image \(\bm x_N\) with \(\sigma_N=0\) at the place where the log probability density is
maximized. 
The optimization objective is to minimize the \(L_2\) denoising error for samples drawn from \(p_{data}\) separately for every \(\sigma\). Defining a neural denoiser \(D(\bm x,\sigma)\):
\begin{equation}
\mathbb{E}_{\bm y\sim p_{data}}\mathbb{E}_{\bm n\sim \mathcal{N}(\textbf{0},\sigma^2\textbf{I})}||D(\bm y+\bm n,\sigma)-\bm y||^2_2,
\end{equation}
where \(\bm y\) is a training image and \(\bm n\) is the sampled Gaussian noise. In this vein, the score of the state with noise of deviation \(\sigma\) can be formulated as:
\begin{equation}
\label{score_func}
\nabla_{\bm x}\log{p(\bm x;\sigma)}=(D(\bm x;\sigma)-\bm x)/\sigma^2.
\end{equation}
For the sampling of the diffusion modeling above, Song et al.~\cite{song2020score} present a stochastic differential equation (SDE) to unify the processes of noise removal and addition into an integral theoretical framework.
An overview of legacy diffusion with a simplified form of SDEs, ordinary differential equations (ODEs), is shown in Fig.~\ref{fig:pipeline}. 
To solve the ODE is to substitute Eq.~\ref{score_func} into it and calculate the numerical integration, i.e., taking finite steps over discrete time intervals.

\subsection{Learned Lossy Image Compression} \label{rate-variable-learned-image-compression}

Learned lossy image compression methods are built upon a variational autoencoder (VAE) framework proposed by Ballé et al.~\cite{balle2018variational}.
The VAE based LIC framework mainly comprises an autoencoder and an entropy model. 
The autoencoder conducts nonlinear transforms between the image space, i.e., input: \(\bm x\), reconstruction: \(\hat {\bm x}\), and the latent representation space, i.e., latent representation: \(\bm y\), quantized latents: \(\hat {\bm y}\); while, the entropy model minimizes the code length by estimating the probability distribution of latent representations. 
\paragraph{Rate-distortion optimization.} In their seminal work, Ballé et al.~\cite{balle2016end} established the end-to-end rate-distortion minimization framework.
It showed that the smallest average code length of latent representation is given by the Shannon cross entropy~\cite{shannon1948mathematical} between the actual marginal distribution and a learned entropy model. 
Thus, the optimization objective appears a rate-distortion trade-off between rate loss \(R(\cdot)\) and distortion loss \(D(\cdot)\):
\begin{equation}
\label{oldRDloss}
\mathcal{L}_{R-D}=\mathcal{R}(\hat {\bm y})+\lambda\cdot\mathcal{D}(\hat {\bm x},\bm x),
\end{equation}
where \(\lambda\) controls whether the network is more concerned about the quality of the recovery or the compression efficiency. 
The optimization problem under the fixed \(\lambda\) in Eq.~\ref{oldRDloss} is employed for the fixed-rate paradigm, driving the encoder to adjust the information reserved in the latent variables $\bm y$.

\paragraph{Quantization scaling.} Nevertheless, the LIC network trained under this schedule only yields a single compression rate result, which limits the design space of diffusion modeling. 
A general solution is to randomly sample \(\lambda\) during training, finetuning a vanilla fixed-rate LIC network to support rate-variable compression~\cite{toderici2015variable,choi2019variable,cui2021asymmetric,wang2023evc}.
In this work, we adopt quantization scaling to control the compression rate via the entropy model. 
Under this circumstance, the role of ``information reducer'' moves from the encoder to the entropy model, separating the autoencoder as a stand-alone component. 
This strategy focuses on the quantization operation \(\lceil \bm y\rfloor\). 
Since codecs only work on integers, the entropy model quantized the latent representation
for bit stream transmission. 
The quantization operation can be regarded as adding a uniform noise in a range of \([-0.5,0.5]\):
\begin{equation}
\label{eq:quant_sim}
\hat {\bm y}=\lceil \bm y\rfloor\Rightarrow \bm y+\mathcal{U}(-0.5,0.5).
\end{equation}
The idea of quantization scaling is to scale the latent representation \(\bm y\) before quantization. Given a scale parameter \(q\), this can be formulated as:
\begin{equation}
\begin{aligned}
\hat{\bm y}_q&=\lceil \bm y/q\rfloor\cdot q\\
		&\Rightarrow [(\bm y/q)+\mathcal{U}(-0.5,0.5)]\cdot q\\
		&=\bm y+\mathcal{U}(-0.5,0.5)\cdot q.
\end{aligned}
\end{equation}
Referring to Eq.~\ref{eq:quant_sim}, quantization scaling can be considered as scaling the uniform noise to control the information gap between original latents \(\bm y\) and quantized latents \(\hat {\bm y}_q\). 
At this point, we obtain a rate-variable compression network with only one parameter to adjust the compression rate. 
This facilitates our following theory construction and its implementation.

\paragraph{Generative compression.} Generative modeling methods, e.g., generative adversarial nets (GANs), vector-quantized variational autoencoders (VQ-VAEs), and diffusion-based models, probabilistically model real data distributions.
Generative compression methods exploit them to supplement the prior distribution to the compressed data, thereby producing photo-realistic reconstructed images.
In their seminal work, GANs and perceptual loss are first utilized to enhance the fidelity of the reconstructed image. 
Agustsson et al.~\cite{agustsson2019generative} first introduced GANs to generate extra details for shaper decompressed images.
Subsequent work has studied further the variants of discriminators, such as patch-GAN~\cite{mentzer2020high}, local label prediction~\cite{muckley2023improving}, and realism guidance~\cite{agustsson2023multi}. 
In the context of the VQ-based method, a vector-quantized variational autoencoder is used to replace or wrap the vanilla LIC variational autoencoder.
Jia et al.~\cite{jia2024generative} move the lossy compression framework into the latent space of VQ-VAEs.
Li et al.~\cite{li2024once} modify the basic VQ-VAEs to obtain a rate-variable generative image compression network.
Given the surprising results of diffusion modeling in the generative domain, recent work has attempted to incorporate the advantages of the diffusion model into image compression. 
The existing diffusion-based methods employ two primary training approaches. 
The first approach involves training a denoising network itself~\cite{yang2023lossy,hoogeboom2023high}, while the second approach involves fine-tuning a pretrained large-scale diffusion model~\cite{careil2023towards,lei2023text+}. 
With respect to the integration of diffusion models within the LIC network, the prevailing methodologies encompass two approaches: the replacement of the decoder in the compression autoencoder with the diffusion model~\cite{yang2023lossy}, and the subsequent addition of the denoising network following the completion of the compression procedure~\cite{hoogeboom2023high}.
Existing diffusion-based methods indirectly take advantage of diffusion modeling.
We expect to establish a novel framework to seamlessly combine LIC and diffusion, realizing the potential of diffusion modeling.
Deriving from the nature of the diffusion modeling is a delicate way.

\section{Method}\label{method}

In the context of learned image compression technologies, the loss of detailed information occurs in the encoder; while the decoder decodes the broken data to images. 
Restricted to the training strategy and network paradigm, it is difficult for the decoder to restore the lost image details. 
We propose generative image compression by estimating
gradients of rate-variable feature distribution to reverse the compression process, assisting the restoration of lost details.
Note that we consider our proposed diffusion modeling specialized in image compression to be regarded
as a form of ``generalized diffusion''. 
Consequently, our analysis lives outside the confines of the diffusion theoretical frameworks but borrows some analytical processes and ideas from these frameworks~\cite{song2019generative,song2020score,karras2022elucidating}. 
Building a standard diffusion model involves two key components, i.e., degraded data construction for training a denoising network (forward process) and the sampling design (reverse process). 
In this section, we construct the diffusion modeling specialized in image compression following the above modules.

\subsection{Compression Forward Process}\label{rate-variable-diffusion-process}

Given an image \(\bm x_0\), the legacy diffusion obtains degraded data \(\bm x_i\) by adding various levels (denoted by deviations \(\sigma_N=\sigma_{max}>\cdots>\sigma_1>\sigma_0=0\)) of Gaussian noise so that \(p(\bm x_i|\bm x_0)\sim\mathcal{N}(\bm x_0,\sigma_i^2\textbf{I})\). 
This corrupts data to varying degrees, and the network is trained to estimate the score function \(\nabla_{\bm x}\log{p(\bm x_i)}\) to restore the original data. 
In essence, learning to restore from corrupted data, also known as the reverse process, enables the network to perform score matching.
The training of the reverse process is relaxed and facilitated by data at various levels of corruption.

For the image compression task, we expect to equip the reverse neural network with the ability to restore the compression-corrupted data. 
Following the standard diffusion process, we replace the data corruption of adding noises with rate-variable compression. 
Defining a pretrained rate-variable entropy model \(E\), the compression process can be formulated as:
\begin{equation}
\label{eq:entropy-model-forward}
\bm x_t=E(\bm x_0,q_t),
\end{equation}
where \(q_t\) denotes the parameter of quantization scaling as we mentioned in Sec~\ref{rate-variable-learned-image-compression}. 
Similar to the deviation \(\sigma\) in legacy diffusion, \(q\) reflects the extent of data corruption.

The reverse neural network \(D_\theta\) inverts the corruption due to compression:
\begin{equation}
\hat{\bm x}_0=D_{\theta}(\bm x_t,q_t),
\end{equation}
where \(\hat {\bm x}_0\) denotes the approximated recovered data produced by network \(D_\theta\). When \(q_t\) is small, \(\hat {\bm x}_0\) should appear close to real \(\bm x_0\) and vice versa. 
Thus, our optimization objective is to minimize the distance between ground truth \(\bm x_0\) and approximated \(\hat {\bm x}_0\):
\begin{equation}
\mathbb{E}_{\bm x_0\sim p_{data},q_t\sim \epsilon_{q}}||\bm x_0-D_{\theta}(E(\bm x_0,q_t),q_t)||^2,
\end{equation}
where \(\bm x_0\) is randomly drawn from a dataset and \(q_t\) is sampled from the distribution \(\epsilon_q\). 
Due to the equal significance of all bit rates in the context of rate-variable compression tasks, we set the
distribution \(\epsilon_q:=\mathcal{U}(q_{min},q_{max})\), where \(q_{max}\) is the maximum supported scale parameter for the entropy
model and \(q_{min}\) is lower than the minimum support, a constant very close to \(0\).
When the sampled \(q_t\) is not supported by the entropy model, referring to Sec.~\ref{rate-variable-learned-image-compression} we take a simulated quantization method as \(\bm x_t=\bm x_0+\mathcal{U}(-0.5,0.5)\cdot q_t\).

\subsection{Reverse Process Design}\label{sampling-design}

In their pioneering work, Song et al.~\cite{song2020score} present a stochastic differential equation (SDE) that maintains the desired distribution \(p\) as the sample \(\bm x\) evolves over time. 
Within this theoretical framework, the sampling process is defined as a reverse SDE, which facilitates the derivation of sampling formulas oriented to better generation results. 
Following this, we endeavor to formulate a sampling approach that aligns with the diffusion modeling of our design for image compression. 
To that end, we first express the aforementioned diffusion process as an ordinary differential equation (ODE). 
Subsequently, a reverse ODE is derived and extended to SDEs for enhancing the quality of the reconstructed images.

\paragraph{ODE formulation.} Although our diffusion forward process is implemented by the entropy model \(E\), we can approximate it (see Sec.~\ref{rate-variable-learned-image-compression}) as follows:
\begin{equation}
\bm x_t\Rightarrow\lceil \bm x_0/q(t)\rfloor\cdot q(t)\Rightarrow \bm x_0+\mathcal{U}(-0.5,0.5)\cdot q(t).
\end{equation}
For the sake of representing differential equations, we move the subscript \(q_t\) into the function brackets \(q(t)\). 
We define the ODE evolving a sample \(\bm x_a\sim p(\bm x_a;q(t_a))\) from time \(t_a\) to \(t_b\) yields a sample \(\bm x_b\sim p(\bm x_b;q(t_b))\), which is satisfied by
\begin{equation}
\label{eq:myode}
\mathrm{d}\bm x=-\frac{\mathrm{d}q(t)}{\mathrm{d}t}\nabla_{\bm x}\log{p(\bm x;q(t))}\mathrm{d}t,
\end{equation}
where \(\nabla_{\bm x}\log{p(\bm x;q(t))}\) is the score function in our theoretical framework. 
To reverse it, we define the reverse neural network $D_\theta$ producing a result \(\hat {\bm 
x}_0\) approximated to \(\bm x_0\).
Thus, the score is obtained by \(\nabla_{\bm x}\log{p(\bm x;q(t))}=(\hat {\bm x}_0-\bm x_t)/q(t)\). 
Following the previous works, the Euler\textquotesingle s method is adopted as the discrete solution during sampling. 
We substitute the equation above to Eq.~\ref{eq:myode} and use Euler\textquotesingle s solver (see Appendix~\ref{odederive}):
\begin{equation}
\bm x_{i+1}=\bm x_i+\frac{q(t_i)-q(t_{i+1})}{q(t_i)}(\hat {\bm x}_0-\bm x_i).
\end{equation}
Nevertheless, the reverse ODE in isolation remains not an optimal solution for sampling, since we find that the deterministic sampling based on the derivation of ODE produces a suboptimal image quality compared to SDE-based stochastic sampling.

\paragraph{Stochastic sampling.} The reverse ODE, corresponding to deterministic sampling, has been observed to result in a worse performance~\cite{song2020denoising,song2020score} than stochastic sampling, i.e., reverse SDEs. 
Following the existing work~\cite{karras2022elucidating}, we extend the ODEs Eq.~\ref{eq:myode} to SDEs and append the random
items from Langevin sampling:
\begin{equation}
\label{eq:SDE}
\mathrm{d}\bm x=-\frac{\mathrm{d}q(t)}{\mathrm{d}t}\nabla_{\bm x}\log{p(\bm x;q(t))}\mathrm{d}t+\alpha(t)\, \mathrm{d}\omega_t\pm\alpha(t)\, \nabla_{\bm x}\log{p(\bm x;q(t))},
\end{equation}
where \(\omega_t\) is the standard Wiener process to inject the randomness to the sampling, the last term is the deterministic decaying item, and \(\alpha(t)\) is the hyperparameter set empirically.
Generally speaking, randomness injection is implemented by adding standard Gaussian noises in the legacy diffusion model framework~\cite{song2019generative}; however, our framework lives outside the standard diffusion theory, so we discuss the form of randomness injection and hyperparameter schedules in the following section.

\paragraph{Randomness injection.} Previous results~\cite{song2020denoising,song2020score} show that finding an optimal setting of stochasticity is significant and that the setting should be empirically determined with respect to specific diffusion models.
In legacy diffusion models, the amount of stochasticity grows with the number of sampling steps.
Recent work~\cite{karras2022elucidating} suggests that a non-uniform growing schedule surpasses a linear one.
Combining the above with the properties of compression tasks, we propose a randomness injection schedule for the hyperparameter $\alpha_t$:
\begin{equation}
\alpha_t=\beta\cdot\sqrt{q_t-q_{min}},
\end{equation}
where $q_t$ adjusts the compression rate, $q_{min}$ is the minimus quantization scale parameter supported by the entropy model, and $\beta$ controls the growth rate of stochastic injections with $q_t$. Moreover, due to the specialization of our model, the specific form of stochasticity is worth arguing, e.g., standard Gaussian noise, uniform noise, and noise drawn from the entropy model. Our analysis in Sec.~\ref{model-analysis} discuss the setting of $\beta$ and the form of randomness injection $\omega$.

\begin{algorithm}[t]
\setstretch{1.1}
\caption{The procedure of our diffusion-based compression of data $\bm x$}
\label{alg:compression}
\DontPrintSemicolon
\setcounter{AlgoLine}{0}
\SetNlSty{}{}{}
\SetNlSkip{-0.8em}
\SetAlgoNlRelativeSize{-1}
\SetKw{AEncoding}{Encoding}
\SetKw{ADecoding}{Decoding}
\SetKw{Asample}{sample}
\SetKw{AGiven}{Given:}
\SetKwComment{Comment}{$\triangleright$\:}{}
\SetKwComment{nopreComment}{}{}
\Indp
\KwIn {$\bm x,q_0$\quad\AGiven$\mathcal{E}(\bm x),\mathcal{D}(\bar{\bm y}),E_\phi(\bm y,q),D_\theta(\hat{\bm y},q),q_{i\in\{0,\cdots,N\}},\alpha,\omega$} 
\AEncoding \nopreComment*{\raisebox{-0.5\baselineskip}[0pt][0pt]{\textnormal{$\triangleright$\:$\bm x$ to be compressed, $q_0$ setting the compression rate}}}
\Indp
    $\bm y_0\leftarrow\mathcal{E}(\bm x)$ \;
    $\hat{\bm y}_q\leftarrow E_\phi(\bm y_0,q_0)$ \Comment*{\textnormal{Approximate $p(\hat{\bm y}_q)$ and compress $\bm y_0$ with scale $q_0$}}
    $\hat{\bm y}_q\xleftrightarrow[]{p(\hat{\bm y}_q)}\text{bit stream}$ \Comment*{\textnormal{Entropy code using $p(\hat{\bm y}_q)$}}
\Indm
\ADecoding \nopreComment*{\raisebox{-0.5\baselineskip}[0pt][0pt]{\textnormal{$\triangleright$\:Reverse directly from $\bar{\bm y}_0:=\hat{\bm y}_q$}}}
\Indp
    \For{$i \in \{0,\cdots,N-1\}$}{ 
        $\bm d_i\leftarrow(D_\theta(\bar{\bm y}_i,q_i)-\bar{\bm y}_i)/q_i$ \Comment*{\textnormal{Evaluate the score $\nabla_{\bm x}\log{p(\bm x;q_i)}$ at $q_i$}}
        $\bar {\bm y}_{i+1}\leftarrow\bar{\bm y}_i+(q_i-q_{i+1})\bm d_i$ \Comment*{\textnormal{Take Euler step from $q_i$ to $q_{i+1}$}}
        \Asample $\bm \epsilon_i\sim \omega$ \nopreComment*{\raisebox{-0.5\baselineskip}[0pt][0pt]{\textnormal{$\triangleright$\:Inject randomness for stochastic sampling}}}
        $\bar{\bm y}_{i+1}\leftarrow\alpha(\bm \epsilon_i-\bm d_i)$ \;
    }
    $\hat{\bm x}\leftarrow\mathcal{D}(\bar{\bm y}_{\scriptscriptstyle N})$ \;
\Indm
\Return $\hat{\bm x}$
\end{algorithm}

\subsection{Image Compression by Estimating
Gradients of Rate-variable Feature Distribution}\label{diffusion-model-in-the-rate-variable-compression-space}

\paragraph{Training the rate-variable compression entropy model.} To alleviate the computational overhead, we move the compression entropy model and the diffusion network into the latent space of the pretrained VAE~\cite{rombach2022high}, where the original image is compressed initially by a factor of four in spatial dimensions. 
We train the rate-variable entropy model \(E_\phi\) via optimizing a rate-distortion loss:
\begin{equation}
\label{RDloss}
\mathcal{L}_{R-D}=-\log p(\hat {\bm z})-\log p(\hat {\bm y})+\lambda\cdot||\hat {\bm y}-\bm y||^2,
\end{equation}
where \(\bm y\) is the latent representation produced by VAE encoder \(\mathcal{E}(\bm x)\), \(\hat {\bm z}\) is the hyperprior latents, and \(\hat {\bm y}\) is quantized by \(E_\phi(\bm y,q)\). 
For multi-rate training, we randomly sample \(\lambda\) and obtain the corresponding \(q\). 
The network architecture of the entropy model is established in accordance with~\cite{han2024causal}, the most recent SOTA work that attains a great trade-off between inference latency and rate-distortion performance.
Further elaborations about such compression paradigms can be found in the literatures~\cite{balle2016end,balle2018variational,han2024causal}.

\paragraph{Training the reverse neural network.} Inspired by the recent work~\cite{karras2022elucidating}, we borrow the denoising U-net architecture from it.
Following Sec.~\ref{rate-variable-diffusion-process}, in every training iteration a quantization scale \(q_t\) is sampled from the uniform distribution \(q_t\sim\mathcal{U}(q_{min},q_{max})\). 
Given a latent representation \(\bm y_0\) extracted from a sampled image \(\bm x\), the compressed latent representation \(\bm y_t\) is obtained by the entropy model \(\bm y_t=E_\phi(\bm y_0,q_t)\).
The reverse neural network \(D_\theta\) is trained using a single \(L_2\) distance:
\begin{equation}
\mathcal{L}_{\text{diff}}=||\bm y_0-D_{\theta}(\bm y_t,q_t)||^2.
\end{equation}
In order to maintain the purity of the theoretical framework, this work incorporates no generative adversarial networks (GANs) or perceptual loss finetuning stage. 
Nonetheless, the capacity to generate photo-realistic images of the proposed approach is evidenced by its
exceptional performance on a range of perceptual metrics (see Experiment Sec.~\ref{comparison_with_SOTA}).

\paragraph{Putting it together.} The whole process of the proposed Algorithm~\ref{alg:compression} can be elaborated as: Given a source image vector \(\bm x\), the autoencoder contains a parametric analysis transform \(\mathcal{E}\) to obtain the latent representation \(\bm y_0\) from \(\bm x\) and a parametric synthesis transform \(\mathcal{D}\) for reconstruction.
\(\bm y\) is then quantized and compressed by the entropy model \(E_\phi\) with a quantization scale of \(q_0\) to latents \(\hat {\bm y}_q\) for storage or transmission. 
When decoding to the reconstructed image \(\hat {\bm x}\), a reverse network $D_\theta$ is utilized to reverse directly from the compressed data \(\hat {\bm y}_q\) for minimal steps. 
The reversed latent variable \(\bar {\bm y}_{\scriptscriptstyle N}\) is then fed to the parametric synthesis transform \(\mathcal{D}\) for reconstruction \(\hat {\bm x}\).

\begin{figure}[t]
    \centering
    \includegraphics[width=1\linewidth]{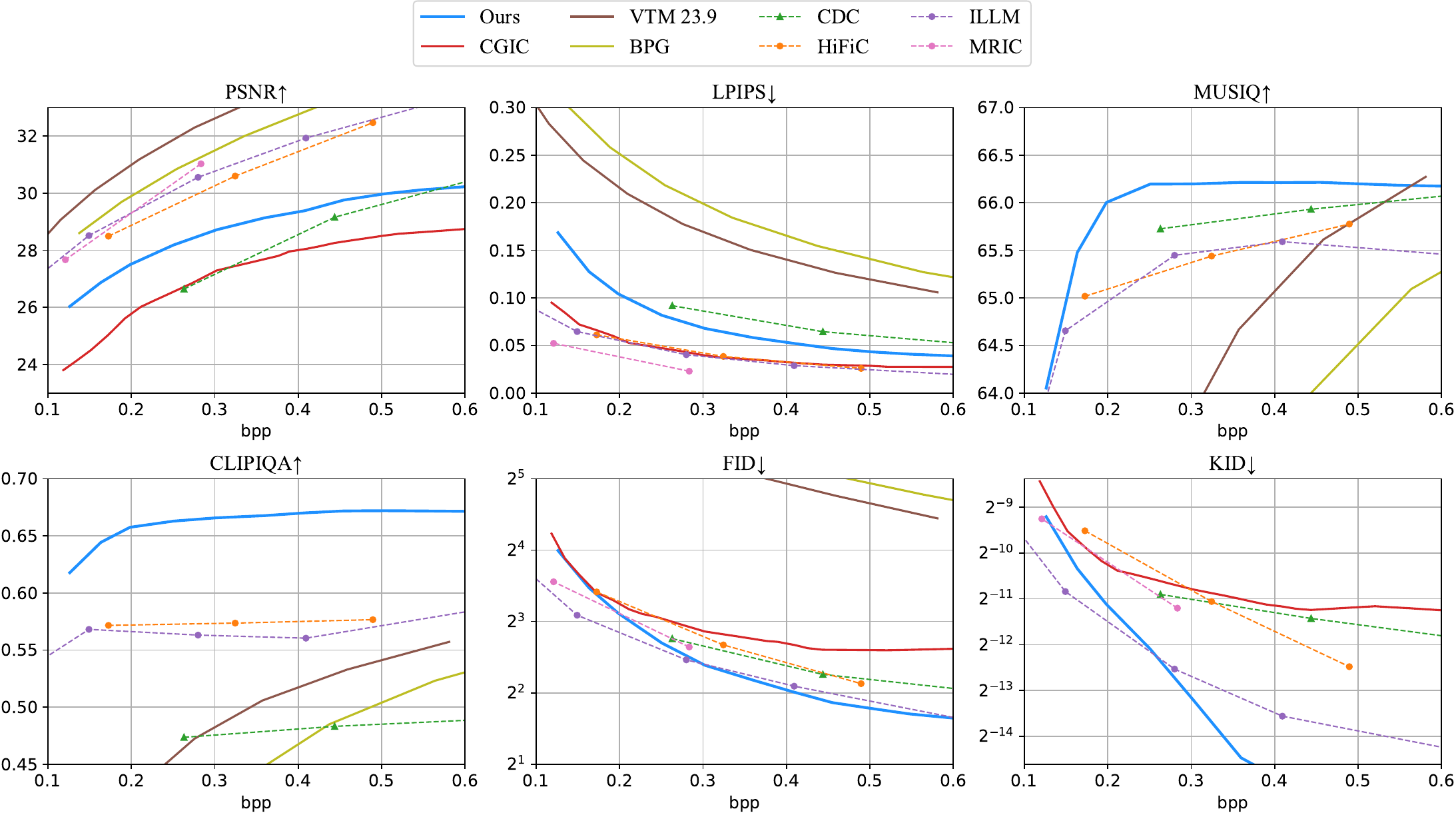}
    \caption{Comparisons of methods across various distortion and statistical fidelity metrics for the DIV2K dataset. The continuous lines represent rate-variable methods (one model for all bit rates). Circular markers denote GAN-based methods and triangular markers denote diffusion-based methods (every marker corresponds to a separate model respectively).}
    \label{fig:RDcurve}
\end{figure}

\section{Experiments}\label{experiments}

\subsection{Experimental Settings}

\paragraph{Datasets.} For the first training stage, we follow the previous compression work~\cite{han2024causal} and train the entropy model on the Open Images~\cite{kuznetsova2020open} dataset.
The randomly selected Open Images dataset contains 300k images with short edge no less than 256 pixels.
For the second training stage, we follow the previous diffusion work~\cite{karras2022elucidating} and train the reverse neural network on the training set of ImageNet~\cite{deng2009Imagenet}.
For evaluation, three benchmarks, i.e., DIV2K dataset~\cite{Agustsson_2017_CVPR_Workshops}, Kodak image set~\cite{kodak}, and CLIC2020 test set~\cite{clic}, are utilized to evaluate the proposed network. 

\paragraph{Implementation details.} The detailed architecture and hyperparameter settings of the entropy model refer to the previous work~\cite{han2024causal}.
For the reverse neural network, we employ the U-Net structure from EDM~\cite{karras2022elucidating}.  
Our experiments and evaluations are carried out on Intel Xeon Platinum 8375C and Nvidia RTX 4090 graphics cards.
By default, our proposed networks are trained using the AdamW optimizer~\cite{loshchilov2017decoupled}.
The weight decay and momenta for AdamW are 0.02 and (0.9, 0.95).
For the entropy model, we randomly crop $256\times256$ sub-blocks from the Open Images dataset~\cite{kuznetsova2020open} with a batchsize of $8$.
We train the entropy model in two stages. In the first (fixed-rate) stage, the model is trained for 0.75M steps using a constant learning rate of $1e-4$. In the second (multi-rate) stage, training continues for another 0.75M steps and then decreases the learning rate to $1e-5$ for 0.375M steps.
The network is optimized with MSE metric, which represents the last term in Eq.~\ref{RDloss}.
For multi-rate training, the multiplier $\lambda$ are $\{0.05,0.025, 0.01, 0.005, 0.001, 0.0005, 0.0001\}$.
As for the reverse neural network, we crop $256\times256$ center blocks from the ImageNet dataset~\cite{deng2009Imagenet} with a batchsize of $128$.
We optimize the network with the initial learning rate $1e-4$ for 0.5M steps and then decrease the learning rate to $5e-5$ for other 0.5M steps.
Since the compression task provides a strong prior (the compressed data), minimal sampling steps are required for the reverse network during the decoding process.
We only use $2$ reverse steps for all bit rates, which improves the efficiency of our method.

\paragraph{Comparison methods and metrics.} We compare our method with the hand-crafted coding standards VVC~\cite{vvc}, BPG~\cite{bpg} and recent state-of-the-art methods~\cite{muckley2023improving,mentzer2020high,yang2023lossy,li2024once,agustsson2023multi}.
Note that our method should be classified with CGIC~\cite{li2024once}, a category of \textbf{rate-variable generative image compression}. 
The rate-variable methods obtain compressed results at all bit rates using only one model, while the other compared methods are fixed-rate, i.e., multiple separate models are required to be retrained for various compression rates.
CDC~\cite{yang2023lossy} is the most recent state-of-the-art \textbf{diffusion-based method}.
Other methods: HiFiC~\cite{mentzer2020high}, MRIC~\cite{agustsson2023multi} and ILLM~\cite{muckley2023improving} are \textbf{GAN-based approaches}.
We use bits per pixel (bpp) value to indicate the compression ratio. 
In addition to the basic distortion metric PSNR, a range of perceptual metrics are used as evaluation: perceptual distortion: LPIPS~\cite{zhang2018unreasonable}, non-reference measure: MUSIQ~\cite{ke2021musiq}, CLIPIQA~\cite{wang2023exploring} and statistical fidelity: FID~\cite{heusel2017gans}, KID~\cite{binkowski2018demystifying}. 
For the calculation of FID and KID, we follow the previous work~\cite{mentzer2020high} to patchify the high-resolution images.

\begin{figure}[t]
    \centering
    \includegraphics[width=1\linewidth]{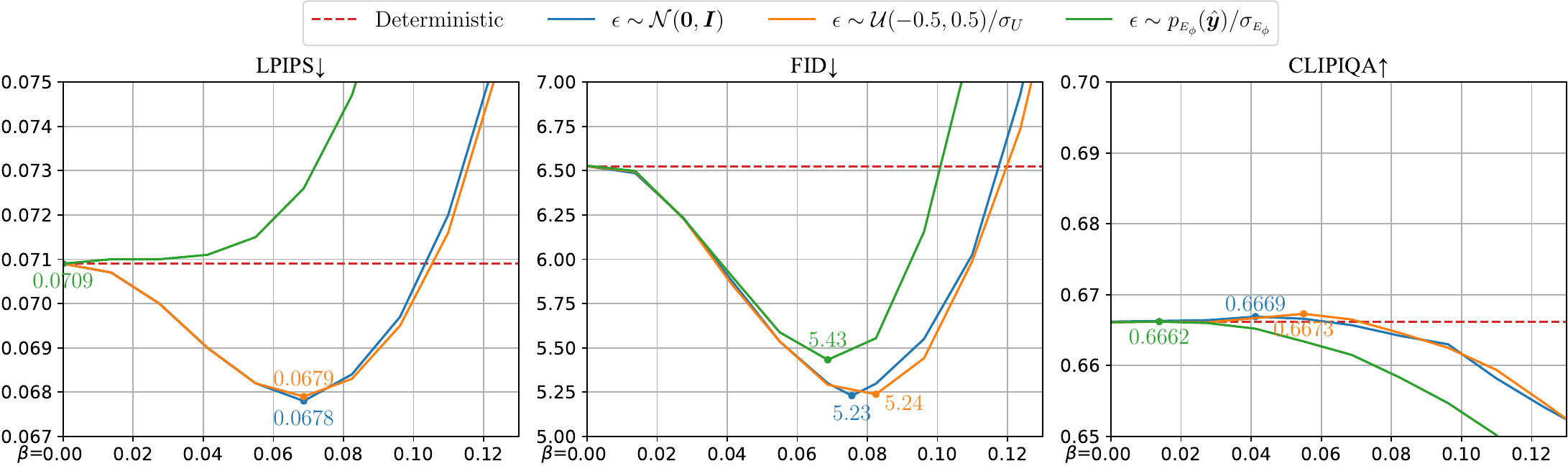}
    \caption{Evaluation of randomness injection schedules when the scale parameter $q_0$ is set as $0.7$ (i.e., bpp $=0.3024$), test on DIV2K with LPIPS, FID, and CLIPIQA. The dashed red lines correspond to deterministic sampling, equivalent to setting $\beta=0$. The blue, orange, and green curves correspond to drawing a noise from a standard normal distribution $\mathcal{N}$, a uniform distribution $\mathcal{U}$, and a probability distribution $p_{\scriptscriptstyle E_\phi}$estimated by the entropy model $E_\phi$, respectively. Note that the latter two distributions are normalized by dividing the statistical deviations $\sigma$. The dots indicate the best observed results.}
    \label{fig:ablation}
\end{figure}

\subsection{Model Analysis}\label{model-analysis}

\paragraph{Form of randomness $\omega$.} We regard three forms of randomness: 
Gaussian noise is the general form adopted in legacy diffusion modeling, uniform noise simulates the quantization operation, and the noise drawn from the entropy model estimated probability distribution of latent variables.
Fig.~\ref{fig:ablation} shows the impacts of various types of randomness. 
We find that standard Gaussian noise and uniform noise are comparable, while, the noise from the entropy model performs worse.
Consequently, we adopt uniform and Gaussian forms of stochasticity as our final randomness injection schedule.


\paragraph{Hyperparameter $\beta$.} We further study the amount of stochasticity.
Following Sec.~\ref{sampling-design}, $\beta$ controls the growth rate of the amount of stochastic injections with $q_t$.
Fig.~\ref{fig:ablation} demonstrates that there exists a trade-off among perceptual distortion (LPIPS), statistical fidelity (FID), and non-reference metric (CLIPIQA).
The results of Gaussian and uniform curves are analogous, and finally we select $\beta=0.075$ as the setting of our SOTA model. 

\begin{figure}
    \centering
    \includegraphics[width=1\linewidth]{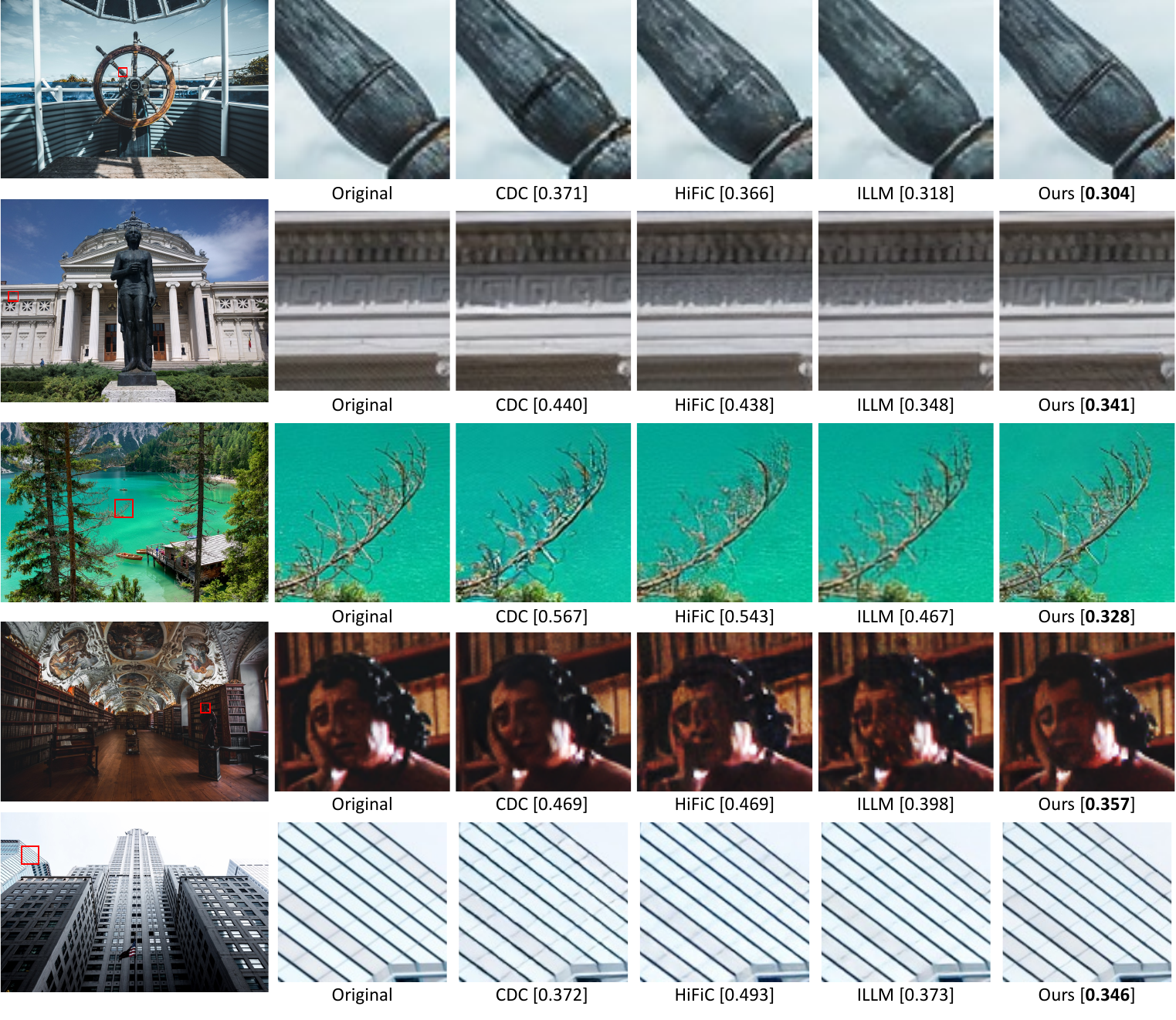}
    \caption{Visualization of the reconstructed images (top to bottom: \textit{0824}, \textit{0812}, \textit{0807}, \textit{0841}, and \textit{0846})
from DIV2K dataset. The titles under the sub-figures are represented as ``method [bpp]''.}
    \vspace{-5pt}
    \label{fig:visual}
\end{figure}

\subsection{Comparison with State-of-the-art Methods}\label{comparison_with_SOTA}

\paragraph{Rate-distortion comparison.} We evaluate the rate-distortion performance of our proposed models by drawing the rate-distortion curves Fig.~\ref{fig:RDcurve}. 
As DIV2K~\cite{Agustsson_2017_CVPR_Workshops} is one of the most commonly used benchmark datasets in the field of low-level vision, we mainly compare our proposed network with the aforementioned SOTA methods on DIV2K dataset. 
Reference models (VTM 23.9 and BPG) achieve the best PSNR scores, but display poor perceptual distortion and statistical fidelity.
GAN-based methods tend to obtain a better LPIPS metric, while, compared to the diffusion-based SOTA method (CDC), ours achieves superior performance.
For the non-reference metrics (MUSIQ and CLIPIQA), our method is able to acquire a clear advantage over the other models.
In the context of rate-variable image compression, our method has uniformly better statistical fidelity than CGIC as measured by FID and KID.
Except for the lower bpp range (below $0.25$), our method performs comparably to the fixed-rate SOTA model ILLM on FID metric, and demonstrates better statistical fidelity evaluated by KID.
We also investigate the effectiveness of our method on the Kodak and CLIC2020 dataset, where the results are provided in our supplementary materials.

\paragraph{Visualization analysis.} Thanks to our well-designed framework, our method achieves superior restoration of fine image details. 
Fig.~\ref{fig:visual} presents five comparison sets against recent state-of-the-art reconstruction models~\cite{yang2023lossy, mentzer2020high, muckley2023improving}, with results generated at comparable bit rates on the DIV2K dataset. 
These visualizations demonstrate that our method faithfully reconstructs image details aligned with the original content, rather than introducing artificial or irrelevant textures.

\section{Conclusion}

In this work, we propose a novel diffusion modeling framework for generative image compression. We establish an organic integration of learned image compression and diffusion, building a complete diffusion framework from the forward process to the reverse process with the help of SDE theory. Our proposed method takes full advantage of the capacity of diffusion modeling, thus achieving state-of-the-art performance on a range of perceptual metrics. Furthermore, we believe this work will spark further innovations across a wide range of domains, especially by encouraging adaptations of the core diffusion modeling framework to adapt to diverse research needs.

\bibliographystyle{splncs04}
\bibliography{egbib}

\begin{thebibliography}{10}
\providecommand{\url}[1]{\texttt{#1}}
\providecommand{\urlprefix}{URL }
\providecommand{\doi}[1]{https://doi.org/#1}

\bibitem{agustsson2023multi}
Agustsson, E., Minnen, D., Toderici, G., Mentzer, F.: Multi-realism image
  compression with a conditional generator. In: Proceedings of the IEEE/CVF
  Conference on Computer Vision and Pattern Recognition. pp. 22324--22333
  (2023)

\bibitem{Agustsson_2017_CVPR_Workshops}
Agustsson, E., Timofte, R.: Ntire 2017 challenge on single image
  super-resolution: Dataset and study. In: The IEEE Conference on Computer
  Vision and Pattern Recognition (CVPR) Workshops (July 2017)

\bibitem{agustsson2019generative}
Agustsson, E., Tschannen, M., Mentzer, F., Timofte, R., Gool, L.V.: Generative
  adversarial networks for extreme learned image compression. In: Proceedings
  of the IEEE/CVF international conference on computer vision. pp. 221--231
  (2019)

\bibitem{balle2016end}
Ball{\'e}, J., Laparra, V., Simoncelli, E.P.: End-to-end optimized image
  compression. arXiv preprint arXiv:1611.01704  (2016)

\bibitem{balle2018variational}
Ball{\'e}, J., Minnen, D., Singh, S., Hwang, S.J., Johnston, N.: Variational
  image compression with a scale hyperprior. arXiv preprint arXiv:1802.01436
  (2018)

\bibitem{bpg}
Bellard, F.: Bpg image format (2015), \url{https://bellard.org/bpg}

\bibitem{binkowski2018demystifying}
Bi{\'n}kowski, M., Sutherland, D.J., Arbel, M., Gretton, A.: Demystifying mmd
  gans. arXiv preprint arXiv:1801.01401  (2018)

\bibitem{careil2023towards}
Careil, M., Muckley, M.J., Verbeek, J., Lathuili{\`e}re, S.: Towards image
  compression with perfect realism at ultra-low bitrates. In: The Twelfth
  International Conference on Learning Representations (2023)

\bibitem{cheng2020learned}
Cheng, Z., Sun, H., Takeuchi, M., Katto, J.: Learned image compression with
  discretized gaussian mixture likelihoods and attention modules. In:
  Proceedings of the IEEE/CVF conference on computer vision and pattern
  recognition. pp. 7939--7948 (2020)

\bibitem{choi2019variable}
Choi, Y., El-Khamy, M., Lee, J.: Variable rate deep image compression with a
  conditional autoencoder. In: Proceedings of the IEEE/CVF international
  conference on computer vision. pp. 3146--3154 (2019)

\bibitem{cui2021asymmetric}
Cui, Z., Wang, J., Gao, S., Guo, T., Feng, Y., Bai, B.: Asymmetric gained deep
  image compression with continuous rate adaptation. In: Proceedings of the
  IEEE/CVF Conference on Computer Vision and Pattern Recognition. pp.
  10532--10541 (2021)

\bibitem{deng2009Imagenet}
Deng, J., Dong, W., Socher, R., Li, L.J., Li, K., Fei-Fei, L.: Imagenet: A
  large-scale hierarchical image database. In: 2009 IEEE conference on computer
  vision and pattern recognition. pp. 248--255. Ieee (2009)

\bibitem{esser2021taming}
Esser, P., Rombach, R., Ommer, B.: Taming transformers for high-resolution
  image synthesis. In: Proceedings of the IEEE/CVF conference on computer
  vision and pattern recognition. pp. 12873--12883 (2021)

\bibitem{goodfellow2014generative}
Goodfellow, I.J., Pouget-Abadie, J., Mirza, M., Xu, B., Warde-Farley, D.,
  Ozair, S., Courville, A., Bengio, Y.: Generative adversarial nets. Advances
  in neural information processing systems  \textbf{27} (2014)

\bibitem{han2024causal}
Han, M., Jiang, S., Li, S., Deng, X., Xu, M., Zhu, C., Gu, S.: Causal context
  adjustment loss for learned image compression. arXiv preprint
  arXiv:2410.04847  (2024)

\bibitem{he2022elic}
He, D., Yang, Z., Peng, W., Ma, R., Qin, H., Wang, Y.: Elic: Efficient learned
  image compression with unevenly grouped space-channel contextual adaptive
  coding. In: Proceedings of the IEEE/CVF Conference on Computer Vision and
  Pattern Recognition. pp. 5718--5727 (2022)

\bibitem{heusel2017gans}
Heusel, M., Ramsauer, H., Unterthiner, T., Nessler, B., Hochreiter, S.: Gans
  trained by a two time-scale update rule converge to a local nash equilibrium.
  Advances in neural information processing systems  \textbf{30} (2017)

\bibitem{ho2020denoising}
Ho, J., Jain, A., Abbeel, P.: Denoising diffusion probabilistic models.
  Advances in neural information processing systems  \textbf{33},  6840--6851
  (2020)

\bibitem{hoogeboom2023high}
Hoogeboom, E., Agustsson, E., Mentzer, F., Versari, L., Toderici, G., Theis,
  L.: High-fidelity image compression with score-based generative models. arXiv
  preprint arXiv:2305.18231  (2023)

\bibitem{jia2024generative}
Jia, Z., Li, J., Li, B., Li, H., Lu, Y.: Generative latent coding for ultra-low
  bitrate image compression. In: Proceedings of the IEEE/CVF Conference on
  Computer Vision and Pattern Recognition. pp. 26088--26098 (2024)

\bibitem{karras2022elucidating}
Karras, T., Aittala, M., Aila, T., Laine, S.: Elucidating the design space of
  diffusion-based generative models. Advances in neural information processing
  systems  \textbf{35},  26565--26577 (2022)

\bibitem{ke2021musiq}
Ke, J., Wang, Q., Wang, Y., Milanfar, P., Yang, F.: Musiq: Multi-scale image
  quality transformer. In: Proceedings of the IEEE/CVF international conference
  on computer vision. pp. 5148--5157 (2021)

\bibitem{kodak}
Kodak, E.: Kodak lossless true color image suite (photocd pcd0992) (1993),
  \url{http://r0k.us/graphics/kodak}

\bibitem{kuznetsova2020open}
Kuznetsova, A., Rom, H., Alldrin, N., Uijlings, J., Krasin, I., Pont-Tuset, J.,
  Kamali, S., Popov, S., Malloci, M., Kolesnikov, A., et~al.: The open images
  dataset v4: Unified image classification, object detection, and visual
  relationship detection at scale. International journal of computer vision
  \textbf{128}(7),  1956--1981 (2020)

\bibitem{lei2023text+}
Lei, E., Uslu, Y.B., Hassani, H., Bidokhti, S.S.: Text+ sketch: Image
  compression at ultra low rates. arXiv preprint arXiv:2307.01944  (2023)

\bibitem{li2024once}
Li, A., Li, F., Liu, Y., Cong, R., Zhao, Y., Bai, H.: Once-for-all:
  Controllable generative image compression with dynamic granularity adaption.
  arXiv preprint arXiv:2406.00758  (2024)

\bibitem{liu2023learned}
Liu, J., Sun, H., Katto, J.: Learned image compression with mixed
  transformer-cnn architectures. In: Proceedings of the IEEE/CVF Conference on
  Computer Vision and Pattern Recognition. pp. 14388--14397 (2023)

\bibitem{loshchilov2017decoupled}
Loshchilov, I., Hutter, F.: Decoupled weight decay regularization. arXiv
  preprint arXiv:1711.05101  (2017)

\bibitem{lu2025learned}
Lu, J., Zhang, L., Zhou, X., Li, M., Li, W., Gu, S.: Learned image compression
  with dictionary-based entropy model. arXiv preprint arXiv:2504.00496  (2025)

\bibitem{mao2024extreme}
Mao, Q., Yang, T., Zhang, Y., Wang, Z., Wang, M., Wang, S., Jin, L., Ma, S.:
  Extreme image compression using fine-tuned vqgans. In: 2024 Data Compression
  Conference (DCC). pp. 203--212. IEEE (2024)

\bibitem{mentzer2020high}
Mentzer, F., Toderici, G.D., Tschannen, M., Agustsson, E.: High-fidelity
  generative image compression. Advances in neural information processing
  systems  \textbf{33},  11913--11924 (2020)

\bibitem{minnen2018joint}
Minnen, D., Ball{\'e}, J., Toderici, G.D.: Joint autoregressive and
  hierarchical priors for learned image compression. Advances in neural
  information processing systems  \textbf{31} (2018)

\bibitem{minnen2020channel}
Minnen, D., Singh, S.: Channel-wise autoregressive entropy models for learned
  image compression. In: 2020 IEEE International Conference on Image Processing
  (ICIP). pp. 3339--3343. IEEE (2020)

\bibitem{muckley2023improving}
Muckley, M.J., El-Nouby, A., Ullrich, K., J{\'e}gou, H., Verbeek, J.: Improving
  statistical fidelity for neural image compression with implicit local
  likelihood models. In: International Conference on Machine Learning. pp.
  25426--25443. PMLR (2023)

\bibitem{qian2022entroformer}
Qian, Y., Lin, M., Sun, X., Tan, Z., Jin, R.: Entroformer: A transformer-based
  entropy model for learned image compression. arXiv preprint arXiv:2202.05492
  (2022)

\bibitem{rombach2022high}
Rombach, R., Blattmann, A., Lorenz, D., Esser, P., Ommer, B.: High-resolution
  image synthesis with latent diffusion models. In: Proceedings of the IEEE/CVF
  conference on computer vision and pattern recognition. pp. 10684--10695
  (2022)

\bibitem{shannon1948mathematical}
Shannon, C.E.: A mathematical theory of communication. The Bell system
  technical journal  \textbf{27}(3),  379--423 (1948)

\bibitem{song2020denoising}
Song, J., Meng, C., Ermon, S.: Denoising diffusion implicit models. arXiv
  preprint arXiv:2010.02502  (2020)

\bibitem{song2019generative}
Song, Y., Ermon, S.: Generative modeling by estimating gradients of the data
  distribution. Advances in neural information processing systems  \textbf{32}
  (2019)

\bibitem{song2020score}
Song, Y., Sohl-Dickstein, J., Kingma, D.P., Kumar, A., Ermon, S., Poole, B.:
  Score-based generative modeling through stochastic differential equations.
  arXiv preprint arXiv:2011.13456  (2020)

\bibitem{vvc}
Team, J.V.E.: Versatile video coding reference software version 23.9(vtm-23.9)
  (2025),
  \url{https://vcgit.hhi.fraunhofer.de/jvet/VVCSoftware_VTM/-/tags/VTM-23.9}

\bibitem{toderici2015variable}
Toderici, G., O'Malley, S.M., Hwang, S.J., Vincent, D., Minnen, D., Baluja, S.,
  Covell, M., Sukthankar, R.: Variable rate image compression with recurrent
  neural networks. arXiv preprint arXiv:1511.06085  (2015)

\bibitem{clic}
Toderici, G., Shi, W., Timofte, R., Theis, L., Ball{\'e}, J., Agustsson, E.,
  Johnston, N., Mentzer, F.: Workshop and challenge on learned image
  compression (clic2020). In: CVPR (2020)

\bibitem{tschannen2018deep}
Tschannen, M., Agustsson, E., Lucic, M.: Deep generative models for
  distribution-preserving lossy compression. Advances in neural information
  processing systems  \textbf{31} (2018)

\bibitem{wang2023evc}
Wang, G.H., Li, J., Li, B., Lu, Y.: Evc: Towards real-time neural image
  compression with mask decay. arXiv preprint arXiv:2302.05071  (2023)

\bibitem{wang2023exploring}
Wang, J., Chan, K.C., Loy, C.C.: Exploring clip for assessing the look and feel
  of images. In: Proceedings of the AAAI conference on artificial intelligence.
  vol.~37, pp. 2555--2563 (2023)

\bibitem{yang2023lossy}
Yang, R., Mandt, S.: Lossy image compression with conditional diffusion models.
  Advances in Neural Information Processing Systems  \textbf{36},  64971--64995
  (2023)

\bibitem{zhang2018unreasonable}
Zhang, R., Isola, P., Efros, A.A., Shechtman, E., Wang, O.: The unreasonable
  effectiveness of deep features as a perceptual metric. In: Proceedings of the
  IEEE conference on computer vision and pattern recognition. pp. 586--595
  (2018)

\end{thebibliography}


\appendix
\newpage

\section{Derivation of Discrete ODE Solver}\label{odederive}

A discrete ODE solver is to use numerical methods to compute integration of ordinary differential equations.
In our framework, Given $\bm x_i$ at a compression ratio of $q_i$, we aim to obtain $\bm x_{i+1}$ at $q_{i+1}$.
Using the first-order Euler\textquotesingle s solver is to exploit a differential approximation:
\begin{equation}
\label{eq:16}
\bm x_{t+\Delta t}=\bm x_t+\Delta t\cdot\frac{\mathrm{d}\bm x}{\mathrm{d}t}.
\end{equation}
Substitute Eq.~\ref{eq:myode} to Eq.~\ref{eq:16}:
\begin{equation}
\bm x_{t+\Delta t}=\bm x_t-\Delta t\cdot\frac{\mathrm{d}q(t)}{\mathrm{d}t}\nabla_{\bm x}\log{p(\bm x;q(t))}.
\end{equation}
We obtain \(\nabla_{\bm x}\log{p(\bm x;q(t))}\) through the neural network $D_{\theta}$:
\begin{equation}
\begin{aligned}
\bm x_{t+\Delta t}&=\bm x_t-\Delta t\cdot\frac{\mathrm{d}q(t)}{\mathrm{d}t}\cdot\frac{\hat{\bm x}_0-\bm x_t}{q(t)},\\ 
&\text{with}\ \hat{\bm x}_0=D_\theta(\bm x_t,q(t)).
\end{aligned}
\end{equation}
To obtain the same form as the main text, we define $t_{i+1}=t_i+\Delta t$:
\begin{equation}
\bm x_{i+1}=\bm x_i-(t_{i+1}-t_i)\cdot\frac{\mathrm{d}q(t)}{\mathrm{d}t}\cdot\frac{\hat{\bm x}_0-\bm x_i}{q(t_i)}.
\end{equation}
For simplicity and continuous sampling during training, we set $q(t):=t$:
\begin{equation}
\bm x_{i+1}=\bm x_i+\frac{q(t_i)-q(t_{i+1})}{q(t_i)}(\hat{\bm x}_0-\bm x_i).
\end{equation}

\section{Compression Latency}

\begin{table}[h]
  \caption{Comparison of coding latency evaluated on Kodak dataset. All the models are evaluated on the same platform. The second line of \textbf{Model} describes the categories of the compared methods.}
  \label{complexity}
  \centering
  \setlength{\tabcolsep}{8pt}
  \begin{tabular}{lcccc}
    \toprule
    \multirow{2}{*}{\textbf{Model}}&Ours&CDC~\cite{yang2023lossy}&ILLM~\cite{muckley2023improving}&CGIC~\cite{li2024once}\\
    \cmidrule{2-5}
    &Diffusion-based&Diffusion-based&GAN-based&VQ-based\\
    \midrule
    \textbf{Encoding Time (ms)}&\centering123&\centering23&\centering60&85\\
    \textbf{Decoding Time (ms)}&\centering280&\centering824&\centering71&32\\
    \textbf{Total Time (ms)}&\centering403&\centering847&\centering131&117\\
  \bottomrule
  \end{tabular}
\end{table}

We compare the coding efficiency of our methods with recent state-of-the-art methods~\cite{yang2023lossy,muckley2023improving,li2024once}. These methods are classified into diffusion-based, GAN-based, and VQ-based approaches. As Table~\ref{complexity} shows, thanks to the minimal sampling steps required for our method, we achieve coding efficiency superior to that of the most recent diffusion-based SOTA work CDC. However, diffusion models generally exhibit slower coding speeds than GAN-based and VQ-based methods. The reason for this is that GAN-based methods only require a one-through transformation, whereas VQ-based methods abandon the entropy model to estimate the probability distribution, which is replaced by transmitting the index of codebook. How to further promote the inference latency of diffusion-based learned image compression methods is worth exploring in the future.

\section{Image Reconstruction Visualization }

\begin{figure}[h]
    \centering
    \includegraphics[width=1\linewidth]{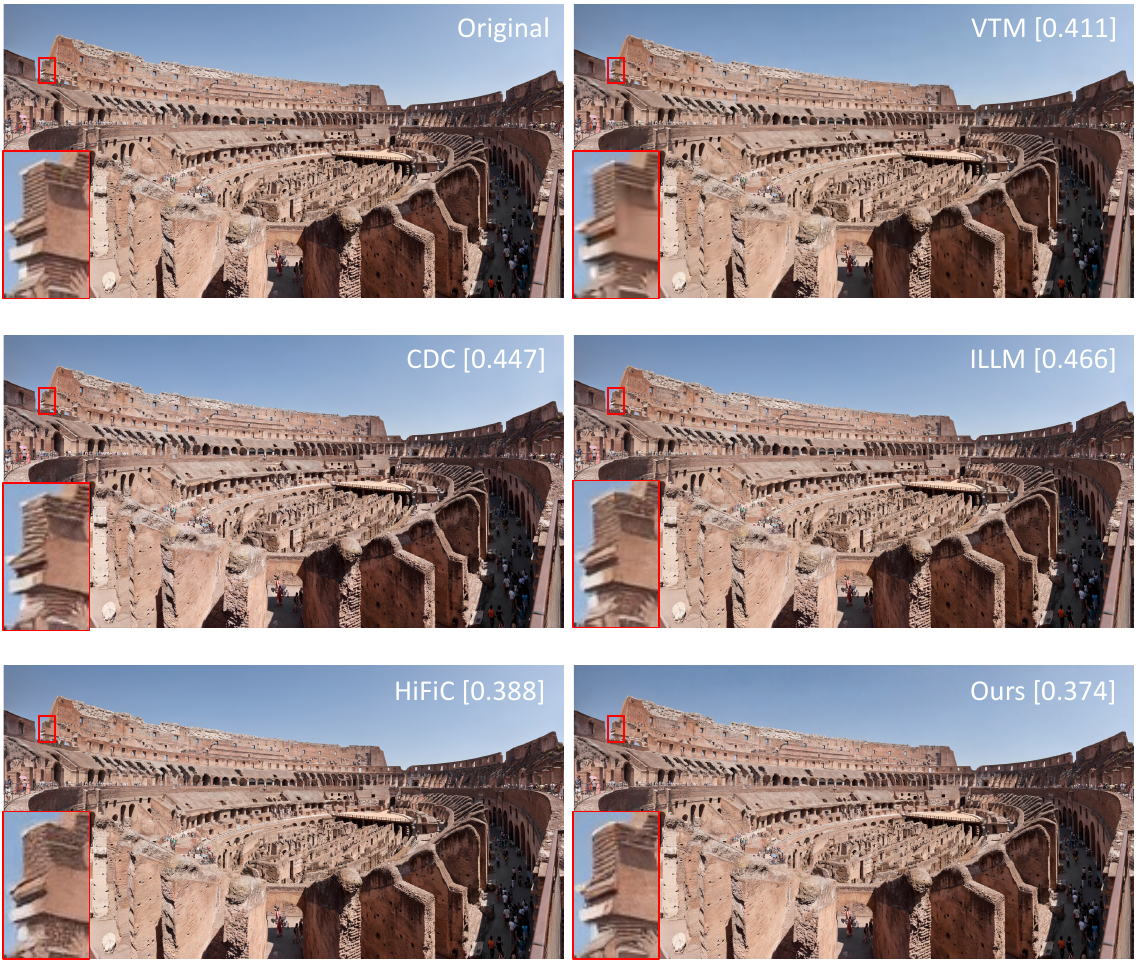}
    \caption{Visualization of the reconstructed images (\textit{0854})
from DIV2K dataset. The titles under the sub-figures are represented as ``method [bpp]''.\\ \makebox[\linewidth][r]{\textit{zoom in for better visualization}}}
    \label{fig:0854}
\end{figure}

\begin{figure}[h]
    \centering
    \includegraphics[width=1\linewidth]{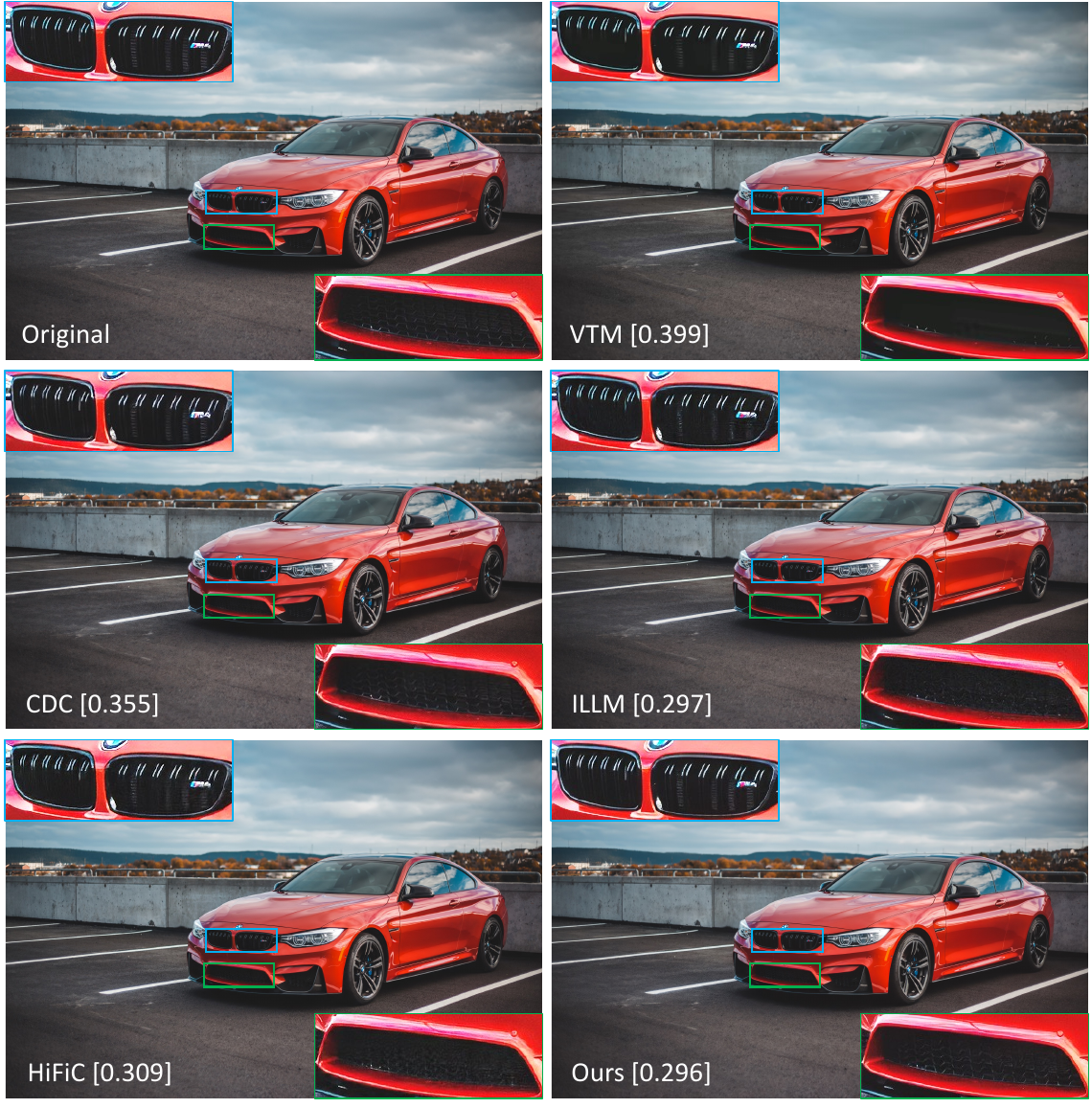}
    \caption{Visualization of the reconstructed images (\textit{1c55})
from CLIC2020 dataset. The titles under the sub-figures are represented as ``method [bpp]''.\\ \makebox[\linewidth][r]{\textit{zoom in for better visualization}}}
    \label{fig:1c55}
\end{figure}

\begin{figure}[h]
    \centering
    \includegraphics[width=1\linewidth]{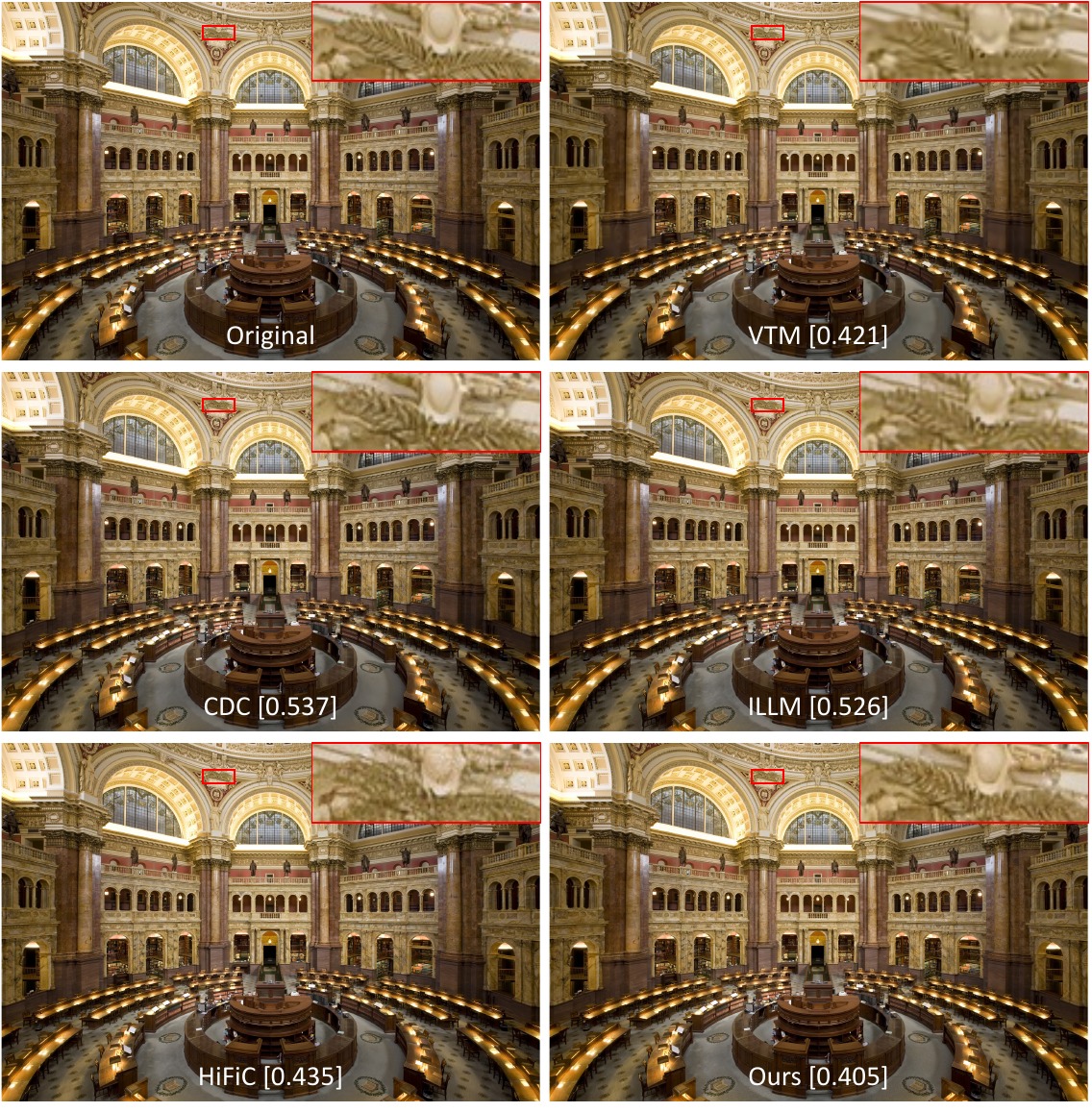}
    \caption{Visualization of the reconstructed images (\textit{0884})
from DIV2K dataset. The titles under the sub-figures are represented as ``method [bpp]''.\\ \makebox[\linewidth][r]{\textit{zoom in for better visualization}}}
    \label{fig:0884}
\end{figure}

We compare the reconstruction results on \textit{0854} (Fig.~\ref{fig:0854}), \textit{1c55} (Fig.~\ref{fig:1c55}), and \textit{0884} (Fig.~\ref{fig:0884}) of our model with those of CDC~\cite{yang2023lossy}, HiFiC~\cite{mentzer2020high}, ILLM~\cite{muckley2023improving} and hand-crafted method VVC~\cite{vvc}.

\vspace*{\fill}
\vspace*{\fill}
\clearpage

\vspace*{\fill}
\vspace*{\fill}
\clearpage

\vspace*{\fill}
\vspace*{\fill}
\clearpage

\section{Further Experimental Results}

\begin{figure}[h]
    \centering
    \includegraphics[width=1\linewidth]{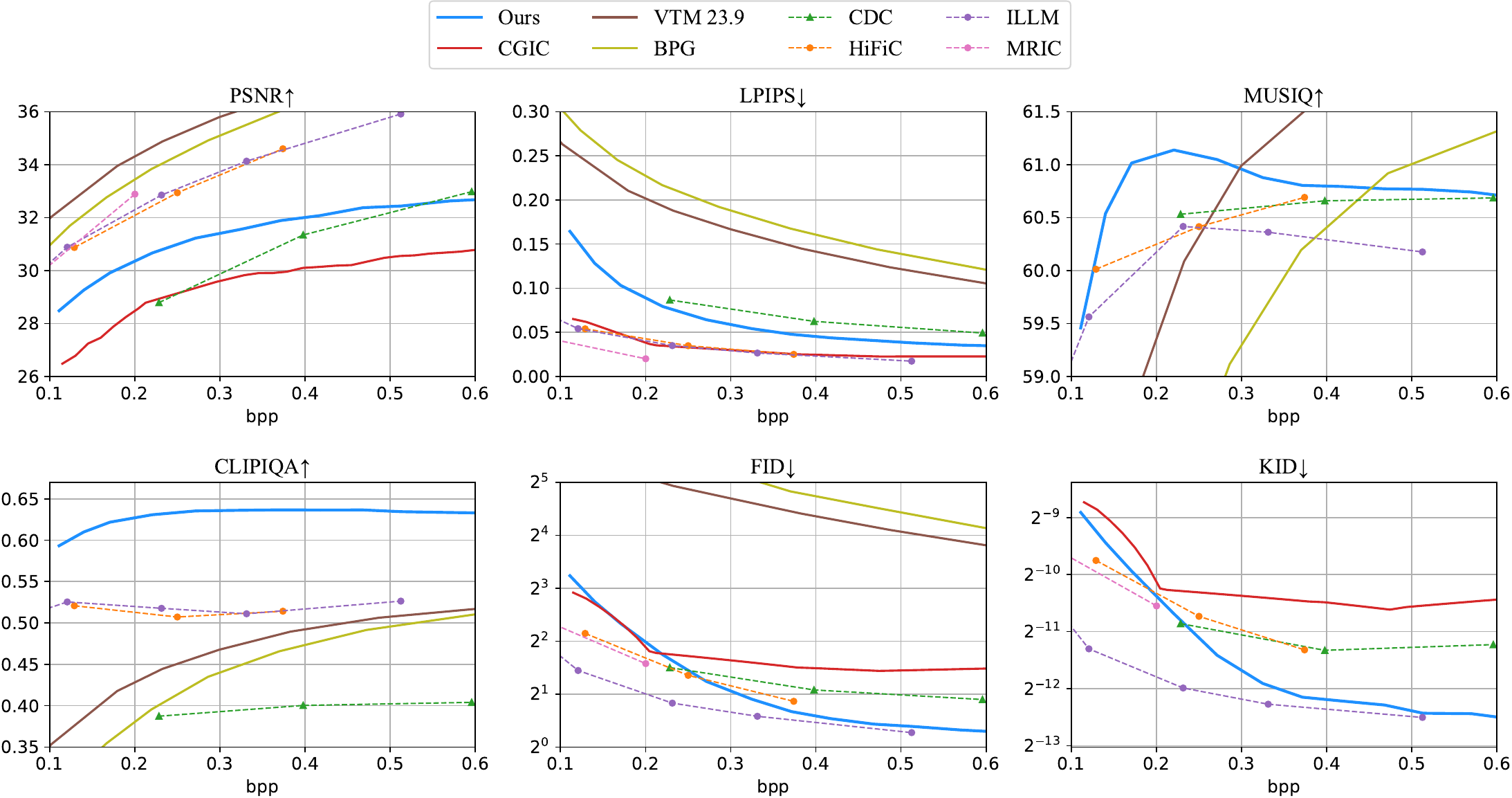}
    \caption{Comparisons of methods across various metrics on the CLIC2020 dataset.}
    \label{fig:clic}
\end{figure}

\begin{figure}[h]
    \centering
    \includegraphics[width=0.6\linewidth]{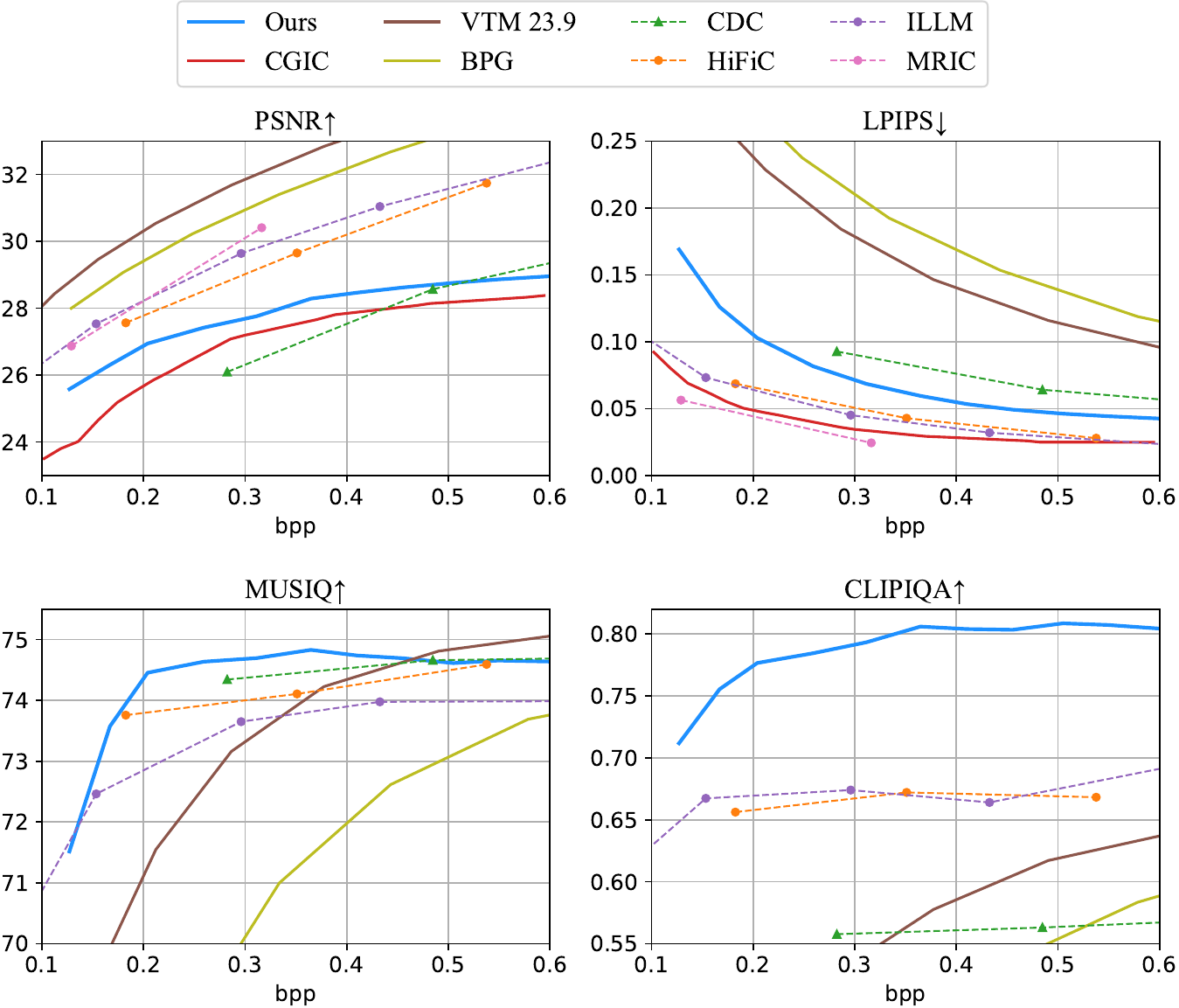}
    \caption{Comparisons of methods across various metrics on the Kodak dataset.}
    \label{fig:kodak}
\end{figure}

We also compare our methods on the CLIC2020 and Kodak datasets. For the Kodak dataset, because too few images are contained, the statistical fidelity metrics (FID and KID) are invalid to evaluate the reconstructed results. The RD curves are revealed in Fig.~\ref{fig:clic} and Fig.~\ref{fig:kodak}.

\vspace*{\fill}
\vspace*{\fill}
\vspace*{\fill}

\end{document}